\documentclass[12pt,preprint]{aastex}

\shorttitle{IMBHs and mass segregation in star clusters}
\shortauthors{Gill et al.}

\begin{document}


\title{Intermediate Mass Black Hole Induced Quenching of Mass Segregation in Star Clusters}


\author{Michael Gill\altaffilmark{1}, Michele Trenti\altaffilmark{2}, M. Coleman Miller\altaffilmark{1}, Roeland van der Marel\altaffilmark{2}, Douglas Hamilton\altaffilmark{1}, Massimo Stiavelli\altaffilmark{2}}

\email{mikegill@astro.umd.edu}

\altaffiltext{1}{University of Maryland, Department of Astronomy, College
  Park, MD 20742-2421 }
\altaffiltext{2}{Space Telescope Science Institute, 3700 San
Martin Drive, Baltimore, MD 21218 USA }


\begin{abstract}
In many theoretical scenarios it is expected that intermediate-mass black
holes (IMBHs, with masses $M\sim 10^{2-4}~M_\odot$) reside at the centers
of some globular clusters.  However, observational evidence for their 
existence is limited.  Several previous numerical
investigations have focused on the impact of an IMBH on the cluster
dynamics or brightness profile.  Here we instead present results from
a large set of direct N-body simulations including single and binary
stars. These show that there is a potentially more detectable IMBH signature,
namely on
the variation of the average stellar mass between the center and the
half-light radius.  We find that the existence of an IMBH quenches mass
segregation and causes the average mass to exhibit only modest radial
variation in collisionally relaxed star clusters. This differs from
when there is no IMBH. To measure this observationally requires
high resolution imaging at the level of that already available from the
Hubble Space Telescope (HST) for the cores of a large sample of galactic
globular clusters. With a modest
additional investment of HST time to acquire fields around the half-light
radius, it will be possible to identify the best candidate clusters 
to harbor an IMBH. This test can be applied only to globulars with a half-light
relaxation time $\lesssim 1$~Gyr, which is required to guarantee
efficient energy equipartition due to two-body relaxation. 
\end{abstract}

\keywords{stellar dynamics --- globular clusters: general --- methods:
  n-body simulations}

\section{Introduction}
Theoretical work has suggested that some globular clusters may harbor
intermediate-mass black holes (IMBHs; $M\sim 10^{2-4}~M_\odot$) in
their centers (e.g., \citealt{por02}). If this is indeed the case, 
there are significant consequences for
ultra-luminous X-ray sources,  gravitational wave emission from dense
star clusters, and the dynamics of globular clusters (GCs) in general 
(see \citealt{vdm04,mc04} for an overview).  Definitive evidence for
IMBHs has, however, been elusive.  For example, \citet{grh02,
grh05} argued for an IMBH in G1 based on the analysis of HST
line-of-sight velocity data and Keck spectra, but an alternative analysis
by \citet{bau03a} points out that acceptable dynamic models without a
large central object also fit the observations. \citet{ger03a,ger03b} 
argued that the kinematics of M15 seem to slightly favor the presence
of an IMBH, but for this cluster alternative interpretations exist
\citep{bau03b,dul03}.  More recently, the observed line-of-sight
kinematics of Omega Cen have also been used to argue for the presence of an IMBH
\citep{ngb08}.

A more secure identification of an IMBH in a GC can, in principle, be
provided by also measuring the proper motion of central stars in order to
reconstruct their orbits and thus firmly establish if a central
massive point object is present.
Several HST-GO programs based on this
idea have been approved in past cycles (e.g. GO10474, PI Drukier;
GO10401 \& GO10841 PI Chandar; GTO/ACS10335 PI Ford), but to date they
have not yielded any indisputable detections. The limitation for such 
studies is the need to carry out multi-year observations, thus progress is
slow. To maximize the chances of success it is thus of primary
importance to focus the observations on the candidates most likely to harbor
an IMBH.

Candidate selection is possible if one focuses on the indirect
influence of the IMBH on the dynamics of its host.  Direct N-body
simulations by \citet{bau04} and \citet{trentiea07b} found that the
presence of an IMBH acts as a central energy source that is able to
prevent gravothermal collapse and thus maintain a sizable core to
half-mass radius ratio throughout the entire life of the GC. The
existence of such a
large ($\gtrsim 0.1$) core to half-mass radius ratio in a collisionally
relaxed cluster might be due to the presence of an IMBH (see also
\citealt{heg07}). However, the picture becomes more complicated when
this signature is transferred from the ideal world of N-body
simulations, where a complete knowledge of the system is available, to
real observations, where essentially only main sequence and red giant
branch stars define the light profile of the system. In fact, an
analysis by \citet{hur07} cautioned that the difference between mass
and light distributions can lead to a large observed core to half-light
radius ratio for GCs with single stars and binaries only.

Here, we continue the search for indirect IMBH fingerprints by
focusing on the consequences of the presence of an IMBH on mass
segregation.  Through direct N-body simulations we show that the
presence of a large (of order 1\% of the total mass), central mass
significantly inhibits the process of mass segregation, even among
only visible main sequence stars and giants.  To the best of our
knowledge this effect was first briefly mentioned in \citet{bau04},
but left without further quantitative analysis. Quenching of mass
segregation is present in all of our simulations with an IMBH,
independent of the initial conditions of the cluster, including
variations in initial mass function, density profile, strength of the
galactic tidal interaction, number of particles and initial binary
fraction.  We find that a differential measurement of the average mass between the
center and the half-light radius is effective in separating star
clusters with and without an IMBH, provided that the stellar system is
at least 5 initial half-mass relaxation times old.  This measure is
observationally feasible with current data (for example, see
\citealt{dem07} and references therein), and can lead to the selection
of a promising set of IMBH host candidates.  A direct observational
application of this approach is left to a companion paper. Here, we focus
instead on building the theoretical framework for such analysis.
In \S~2 we describe our numerical simulations, in \S~3 we discuss our
results, and in \S~4 we present our conclusions.

\section{Numerical Simulations}\label{sec:ns}

The numerical simulations presented in this paper have been carried
out with a state-of-the-art direct N-body code for star cluster
dynamics, NBODY6 \citep{aar03}. NBODY6 has
been modified as discussed in \citet{trentiea07b} to improve accuracy
in the presence of an IMBH, and 
uses regularization of close gravitational encounters without any
softening.  This makes it optimal to follow interactions
within the sphere of influence of the IMBH. 

\subsection{Units and timescales}

NBODY6 uses the standard system of units of
\citet{hm86} in which $G=M=-4E_T=1$, where G is the gravitational
constant, M is the total mass and $E_T$ is the total energy of the
system. In this system of units, the half-mass relaxation time, which
is the relevant timescale for mass segregation and energy
equipartition is defined as follows \citep{spi87}:
\begin{equation}
t_{rh} = \frac{0.138 N  r_h^{3/2}}{\ln{(0.11 N)}}, 
\end{equation}
where $N$ is the number of stars in the system and $r_h$ is the
half-mass radius. In physical units $t_{rh}$ can be expressed as
\citep{djo93}:
\begin{equation}
t_{rh} = \frac{8.9 \cdot 10^5 yr}{\ln(0.11N)} \times \left (
 \frac{1 M_{\sun}}{\langle m_*\rangle} \right )\times \left( \frac{M}{ 1 M_{\sun}} \right )^{0.5} \times 
\left( \frac{r_h}{ 1 pc} \right )^{1.5},
\end{equation}
where $\langle m_* \rangle $ is the average mass of a star.

\subsection{Initial conditions}

Galactic GCs are made of some of the oldest stars in
our galaxy \citep[e.g., see ][]{krau03} and are collisionally relaxed systems.
Their two-body half-mass relaxation times $t_{rh}$ are shorter than
their age \citep[e.g., see][]{heg03}, thus their initial conditions are
largely unknown. However, the evolution on a relaxation timescale is
only weakly dependent on the initial configuration as the system
evolves toward a self-similar configuration where the density and
light profiles are determined primarily by the efficiency of kinetic
energy production in the core due to gravitational encounters
\citep{ves94,trentiea07a,trentiea07b}. In this paper, we explore a
number of different initial configurations, varying the initial mass
function, the initial density profile, the strength of the galactic
tidal field, and the fraction of primordial binaries to verify that the
evolution of the system is indeed independent of the initial configuration. 
The initial density profile is always that of a single-mass 
\citet{kin66} model, but we use a full mass spectrum in the N-body 
calculations. The number of particles is varied 
from $N=8192$ to $N=32768$ to quantify the
evolution of mass segregation with and without an IMBH. The details of
our runs are reported in Table~\ref{tab:sim}. 

Initial stellar masses were drawn from either a \citet{sal55} or
\citet{ms79} initial mass function (IMF, hereafter), that is:
\begin{equation}
\xi(m) \propto m^{\alpha},
\end{equation}
with $\alpha=-2.35$ and $m \in [0.2:100] M_{\sun}$ for the Salpeter
IMF, while for the Miller \& Scalo IMF the power law slope is the
following: $\alpha = -1.25$ for $m \in [0.2:1] M_{\sun}$, $\alpha =
-2.0$ for $m \in [1:2] M_{\sun}$, $\alpha = -2.3$ for $m \in [2:10]
M_{\sun}$, $\alpha = -3.3$ for $m \in [10:100] M_{\sun}$. In addition,
we have also carried out control runs that extend the IMF down to $0.1
M_{\sun}$, as such stars exist in GCs, but are not bright enough 
to be detected in most of the cluster.

We handled stellar evolution by assuming a turnoff mass $M_{T.O.}  =
0.8 M_{\sun}$, and instantaneously reducing all stars to their final
state at the beginning of the simulation. Stars with masses $0.8
M_{\odot} \leq m < 8.0 M_{\odot}$ were assumed to become white dwarfs
and reduced to a final mass as prescribed in \citet{hur00}.  Stars in
the ranges 8.0-25.0$M_{\odot}$ and 25.0-100.0$M_{\odot}$ became
neutron stars (unity retention fraction assumed) and black holes, 
and were reduced linearly to 1.3-2.0$M_{\odot}$ and 5-10$M_{\odot}$, 
respectively.  Our model makes the approximation that most of the 
relevant stellar evolution occurs on a timescale shorter than a 
relaxation time.  This choice is appropriate to model the dynamics 
of an old GC on a relaxation timescale with only a limited number 
of particles and is more realistic than using an unevolved mass 
spectrum appropriate for young star clusters when $N \lesssim 
30000$ \citep{tre08}.

In addition, about half of our runs included primordial binaries, an
important component of many GCs (e.g., see
\citealt{pul03}) that can influence the evolution of mass
segregation. In fact, binaries are on average twice as massive as
singles and thus tend to segregate in the core of the system 
(e.g., see \citealt{htt06}). We define the fraction of binaries to be
\begin{equation}
f_b = n_b / (n_s + n_b)
\end{equation}
where $n_s$ and $n_b$ are the initial number of single stars and binaries,
respectively.  Thus, a run with $N = n_s + n_b = 8192$ and $f_b
= 0.1$ actually has $8192 + 819 = 9011$ objects.  As the dynamical
influence of binaries tends to saturate for $f_b \approx 0.1$
\citep{ves94,htt06}, all our runs with primordial binaries have
$f_b=0.1$, a number similar to the observed binary fraction of many
old GCs (e.g., \citealt{alb01}). Binaries were initialized as in \citet{htt06},
i.e. from a flat distribution in binding energy from
$\epsilon_{min}$ to $133 \epsilon_{min}$, with $\epsilon_{min} =
\langle m_* \rangle \sigma_c(0)$. Here, $\sigma_c(0)$ is the initial
central velocity dispersion of the cluster.

To half of the simulations, we added an IMBH with mass $M_{IMBH}\approx 0.01$ ($\approx 1
\%$ of the entire cluster), the same ratio as a
$\sim10^3M_{\odot}$ black hole would have to a GC of
mass $10^5M_{\odot}$. In some of the simulations (see
Tab.~\ref{tab:sim}), we increased the mass of the IMBH to study the
dependence of mass segregation on this parameter.

All objects in our runs were treated as point masses, thus neglecting
stellar evolution and collisions as well as any growth of the IMBH due to accretion
of tidally disrupted stars.  These effects have only a minor influence
on the late-time dynamics of the cluster, since actual collisions are
rare after massive stars have evolved, and accretion onto the IMBH is
minimal (e.g. see \citealt{bau04}).

The evolution of the clusters includes the tidal force from the parent
galaxy, assuming circular orbits with radii such that the tidal cut-off
radius is self-consistent with the value of the King parameter $W_0$
used. The galactic tidal field is treated as that due to a point mass,
and the tidal force acting on each particle is computed using a linear
approximation of the field. Particles that become unbound are removed
from the system. For full details of the tidal field treatment see
\citet{trentiea07a}.

For validation purposes, we also analyzed a few snapshots from three
runs with $N=131072$ carried out by \citet{baumak03} and by
\citet{bau04} with and without a central IMBH, kindly made available
by the authors. These runs include full
stellar evolution using the \citet{hur00} tracks, but no primordial
binaries. In addition, accretion of tidally disrupted stars onto the IMBH
is included. The initial star positions and velocities from these runs
were also drawn from a King model with $W_0 = 7.0$, but instead
the mass spectrum was drawn from a \citet{krou01} IMF, with
$\alpha=-1.3$ for $m \in [0.1:0.5] M_{\sun}$ and $\alpha=-2.3$ for
$m>0.5M_{\sun}$. The upper cut-off mass was $15 M_{\odot}$ for run
128kk.1, $30 M_{\odot}$ for run 128kkbh.1a, and $100 M_{\odot}$ for run
128kkbh.1b. These snapshots provide us with a control group
against which we can test the validity of our own models, and also allow us
to probe the extrapolation of our results to higher N.

\section{Results}\label{sec:results}

\subsection{Overall evolution of the star clusters}

\begin{figure}[ht]
\plotone{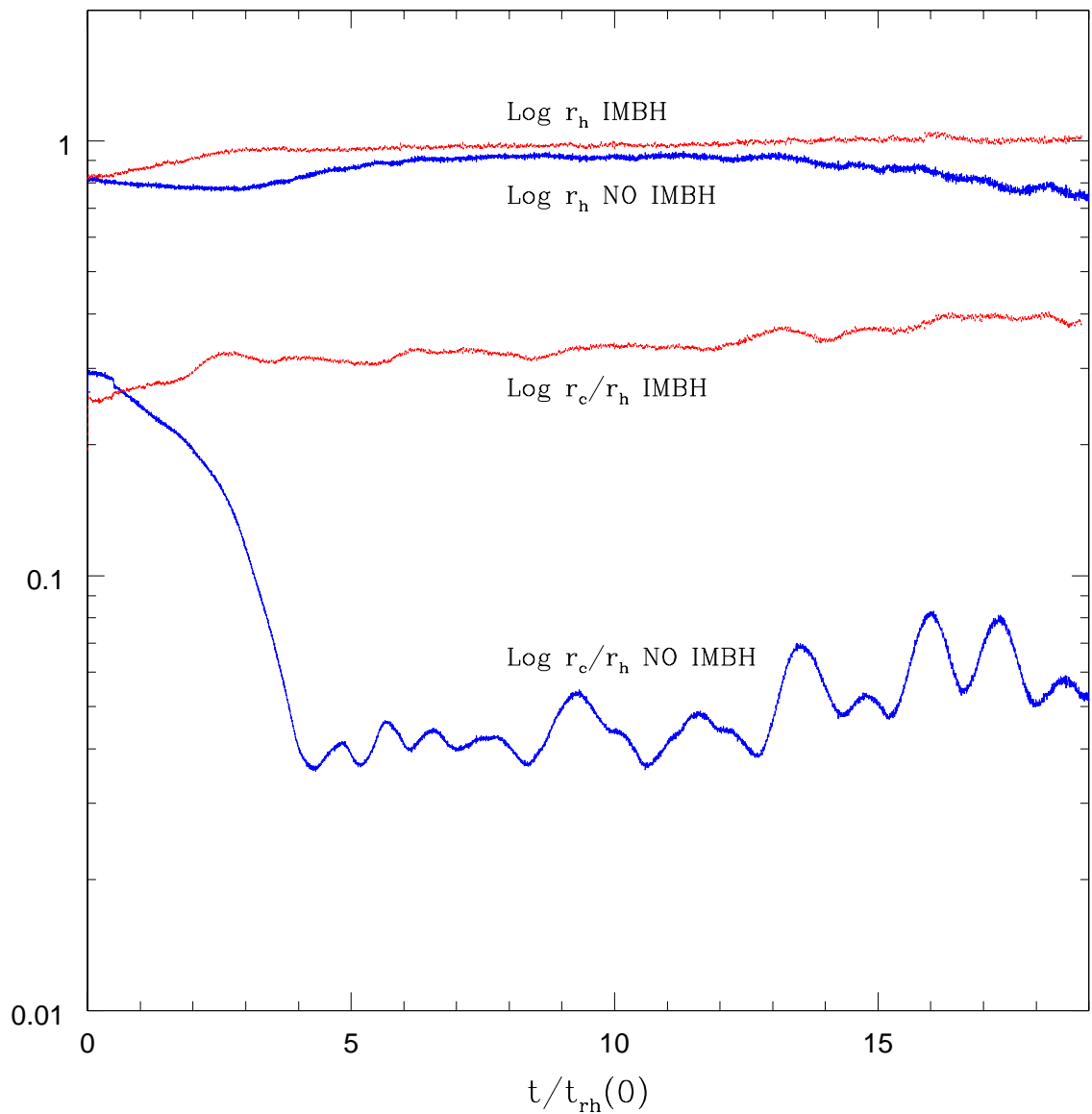}
\caption{Evolution of the three dimensional half mass radius ($r_h$)
and of the core to half mass radius ratio ($r_c/r_h$) in NBODY units for our $N=32769$
simulations with (red curves) and without an IMBH (blue
curves). The presence of an IMBH prevents core collapse.  We have smoothed the
curves by applying a triangular smoothing window of size $1.0 t_{rh}(0)$.}
\label{fig:core}
\end{figure}

A star cluster with single stars only and no central IMBH evolves
toward core collapse within a few relaxation times. The collapse is
eventually halted when the central density is high enough to
dynamically form binaries. At this stage gravothermal oscillations set
in and the density profile of the cluster remains self-similar until
the final stages of tidal dissolution of the system. The existence of
either primordial binaries or an IMBH serves as an energy source to
counteract the collapse, resulting in a more significant core
\citep{trentiea07b}. This is confirmed in all models (for example, see
Fig.~\ref{fig:core} for the evolution of the core and half mass radius
in our 32k simulations). Here, we use the \citet{ch85} definition of the core
radius, namely
\begin{equation}
r_c = \frac{\sum_{i=1}^{N}r_i \rho_i m_i}{\sum_{i=1}^{N}\rho_i m_i},
\end{equation}
where $m_i$ is the mass of the \emph{i}th star, $r_i$ is its from the 
cluster center of mass, and the density $\rho_i$ around each particle 
is calculated using the distance to the fifth closest neighbor.
  
As expected from previous investigations
based on equal mass particles \citep{trentiea07a}, the density profile
of our clusters progresses to a self-similar configuration, which is
independent of the initial configuration of the stars and the
IMF. This independence justifies our treatment of stellar evolution at
the beginning of the simulations, and provides further evidence of the
erasure of initial conditions after a few relaxation times. The
overall evolution of star clusters with and without an IMBH and/or
primordial binaries has been discussed in the literature
\citep{baumak03,bau04,htt06,trentiea07a,trentiea07b,fre07,hur07}. Here
we focus instead on a novel aspect that has a promising observational
signature, namely the evolution of mass segregation for runs with an
IMBH.

\subsection{Mass segregation}

As our overarching goal is to propose a viable observational test to
identify a star cluster that is likely to harbor an IMBH, we took
steps throughout our analysis to replicate observational data as closely
as possible.  We limited our analysis to data projected onto two
dimensions, and excluded stellar remnants from our calculations of
observationally accessible quantities. Most of our runs did not have stars
with masses below $0.2 M_{\odot}$, because these are generally too dim to 
detect with a high completeness. However, we did perform two 
16k control runs with masses down to $0.1 M_{\odot}$ to ensure that the 
presence of smaller, undetectable stars did not affect the mass segregation. 
For a proper comparison to the other runs, 
we did exclude the stars with masses 
between $0.1$ and $0.2 M_{\odot}$ from the calculation of the 
observationally accessible quantities for these runs.

Binary systems were handled by only including the brighter member in the 
analysis of the observationally accessible quantities.
This choice is motivated by the fact that for real observations,
masses will need to be estimated from luminosities.  Since binaries in
GCs are typically not resolved (the separations are below a few astronomical 
units for the range of binding energies considered), 
we observe mainly single sources.
Because the luminosity of a main sequence star is highly sensitive to
its mass ($L \sim M^{3.5}$), the lighter star contributes very little
to the overall luminosity in many cases, and thus the total luminosity
will be very similar to that of the brighter member. 

To quantify the effects of mass segregation, we examined the radial
variations in average stellar mass --- or equivalently --- in the slope of
the mass function (if the mass function is a power law in the mass
range considered, then there is a one-to-one relation between the average
mass and the slope). As a consequence of energy equipartition, heavier
particles sink to the center of the cluster within a few relaxation
times, increasing the difference in average mass between the center
and the halo of the cluster. As our main diagnostic of mass
segregation, we define
\begin{equation}
\Delta \langle {m} \rangle = \langle m \rangle \ (r = 0) - \langle m \rangle (r = r_{hm})
\end{equation}
where $r_{hm}$ is the projected half-mass radius of the cluster
(computed using only visible stars), and $\langle m \rangle$ is the
average mass for main sequence stars with $m \in [0.2:0.8]M_{\sun}$.
Both of these measurements are taken from projected radial bins each
containing 5\% of the cluster's visible stars.  Because nearly all of
the deviation in $\langle m \rangle(r)$ occurs within this radial range, we
are maximizing our baseline for measuring mass segregation while using
fields with a reasonable number of stars. This definition also allows
for a straightforward comparison to observational data as only two
fields per cluster are sufficient.

\begin{figure}[ht]
\plotone{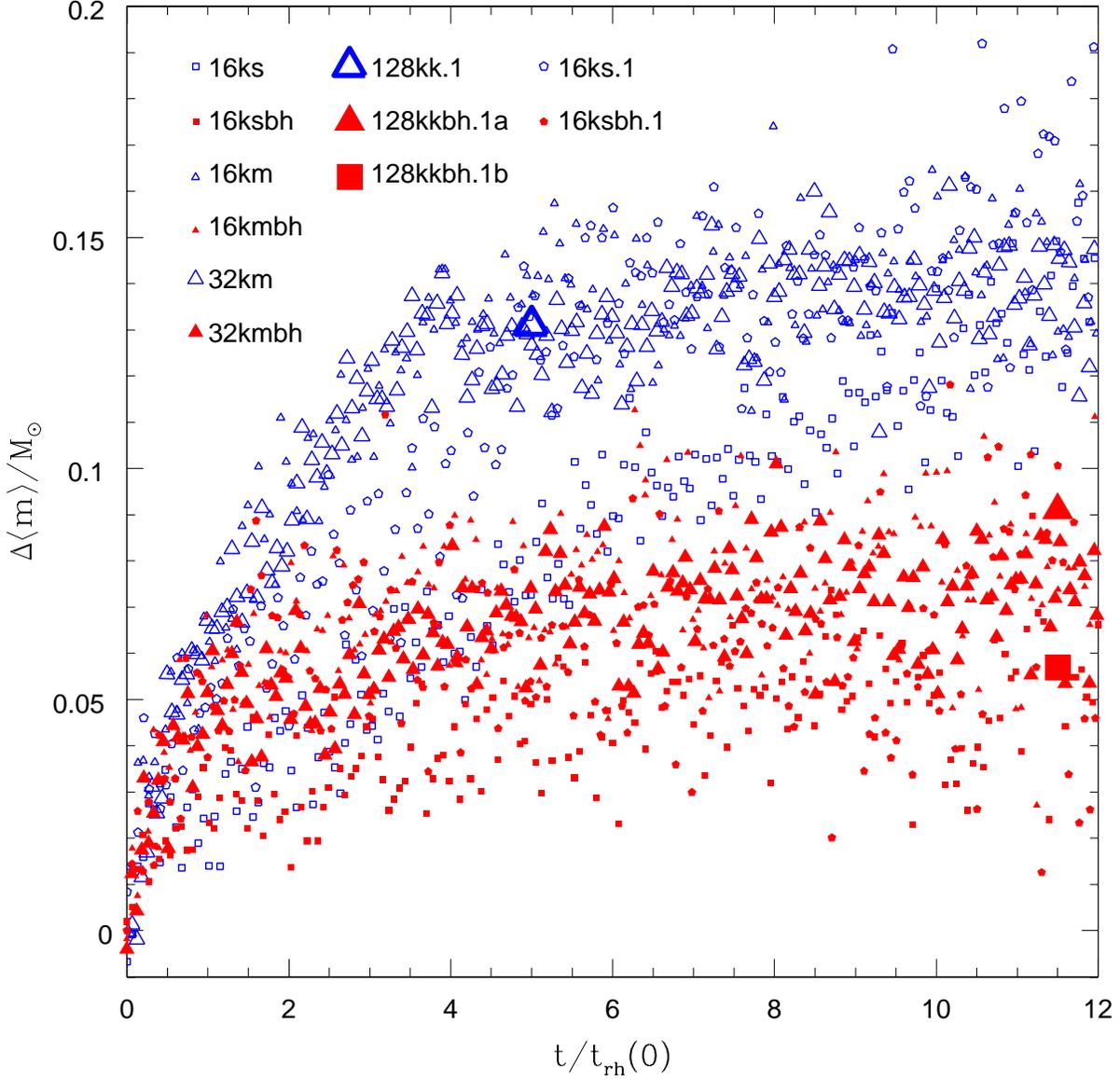}
\caption{Evolution of mass segregation (via $\Delta\langle m \rangle $, expressed in $M_{\sun}$)
  across the span of all N-body simulations with $N \geq 16384$ and
  $f_b = 0$.  Red points are from simulations with an IMBH, while
  blue points represent runs with no massive central object.  The runs
  have no primordial mass segregation ($\Delta\langle m \rangle  = 0$), but
  on a relaxation timescale, the systems settle to a quasi-equilibrium
  configuration with varying degrees of mass segregation.  A central
  IMBH quenches the mass segregation and keeps $\Delta\langle m \rangle 
  \lesssim 0.09 M_{\sun}$.}
\label{fig:16k}
\end{figure}

Fig.~\ref{fig:16k} depicts the evolution of $\Delta\langle m \rangle $ for
our $N=16384$ to $N=32768$ runs without primordial binaries. For each
run, we analyze the configuration of the system every $15$ Nbody units
(which corresponds to more than 10 measurements per relaxation time).
Runs with an IMBH are represented as red points, while runs without
are blue points.  Because they were drawn from single-mass King
models, our clusters begin out of equilibrium.  On a relaxation
timescale, we see them evolve towards a new quasi-equilibrium state.  
After $\sim 5 t_{rh}(0)$, most clusters have settled into this equilibrium, with
those harboring an IMBH showing a smaller amount of
mass segregation, i.e. smaller values of $\Delta\langle m \rangle $. The
points from the control snapshots 128kk.1, 128kkbh.1a and 128kkbh.1b, as well as the 
two 16k control runs (16ks.1 and 16ksbh.1) are also
plotted in Fig.~\ref{fig:16k}, and are in good agreement with those from our models. 
The data in Fig.~\ref{fig:16k} come from a variety
of initial configurations, not only in terms of the particle number
but also in terms of initial mass function. The use of a differential
indicator for mass segregation allows us to cancel out the dependence
on the global value of $\langle m \rangle$ (or on the global mass function
slope).

Simulations with a Salpeter IMF (16ks, 16ksbh) contain many more
massive remnants than the number allowed by a Miller \& Scalo IMF
(16km, 16kmbh, 32km, 32kmbh).  For example, a 16k simulation with a Salpeter IMF
and $m \in [0.2:100]M_{\odot}$ initially contains $\sim20$ stellar mass black holes, whereas a Miller \&
Scalo IMF will only have $\sim1$.  This difference causes us to
observe a slower growth of $\Delta\langle m \rangle $ in the Salpeter IMF
runs, as a central cluster of stellar mass black holes partially
quenches mass segregation of visible stars, much like an
IMBH. However, stellar mass black holes eject each other from the
system within a few relaxation times \citep[see also][]{mer04}, so
eventually these systems fully develop the amount of mass segregation
observed in runs starting from a Miller \& Scalo IMF. 

The control runs also reflect this trend.  The 128k run without an IMBH (128kk.1) has a maximum
allowed IMF mass of $15 M_{\odot}$, meaning there are essentially no
massive remnants.  After 5 relaxation times, it is in good agreement
with our simulations drawn from a Miller \& Scalo IMF, which also
produces very few massive remnants. The runs with an IMBH (128kkbh.1a and 128kkbh.1b), which have
maximum initial masses of $30 M_{\odot}$, and $100 M_{\odot}$, respectively,
are both consistent with our other runs with an IMBH.  However,
128kkbh.1a's lower maximum mass results in fewer massive remnants, and
thus a value for $\Delta\langle m \rangle $ closer to 16kmbh, which also
contains few stellar mass black holes.  128kkbh.1b has more massive
remnants and behaves similarly to 16ksbh, as we would expect.  Finally, we see that 
our 16k control runs with a lower IMF cut-off at $0.1 M_{\odot}$
fall somewhere between the Miller \& Scalo runs and the 
Salpeter runs as far as mass segregation is concerned.  Although we draw 
from a Salpeter IMF in the control runs, the lower minimum IMF mass
in these simulations results in fewer massive remnants than the other Salpeter
runs ($\sim8$ stellar mass black holes instead of $\sim20$), but more than a Miller \& Scalo
IMF would produce.  We also note the increased 
scatter in these two runs as a result of excluding the large number of main sequence
stars with masses $0.1 M_{\odot} < m < 0.2 M_{\odot}$
in the calculation of the observationally accessible quantities.  

\begin{figure}[ht]
\plotone{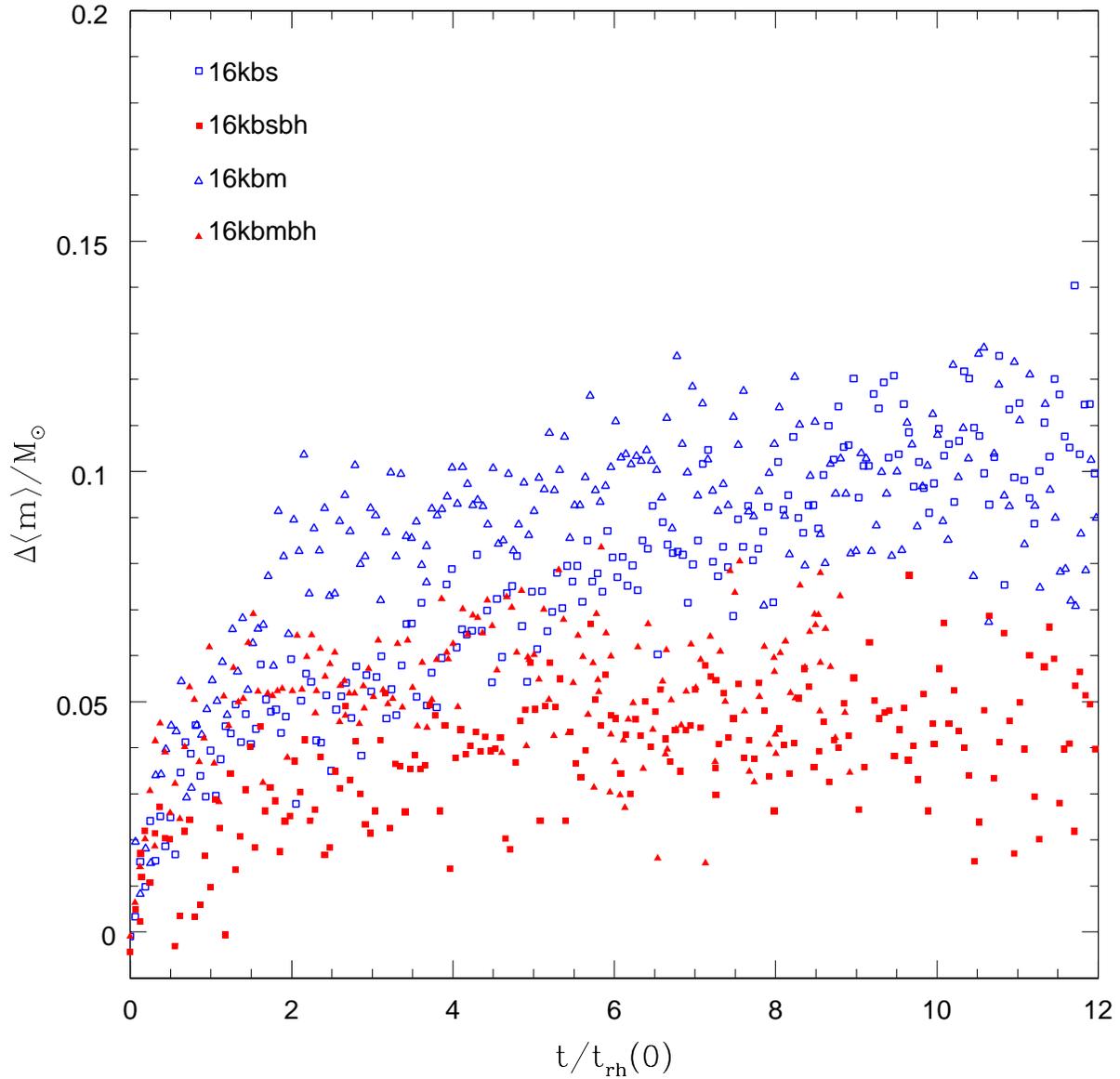}
\caption{Evolution of mass segregation as in Fig. 1, but for our $N
  \geq 16384$ simulations {\it with} primordial binaries.  Qualitatively, we
  see that the results are similar to those of the runs with single
  stars, but the equilibrium values of $\Delta\langle m \rangle $ are
  marginally lower at later times when compared to those where $f_b = 0$.}
\label{fig:16kbin}
\end{figure}

The situation is very similar when primordial binaries are included (see
Fig.~\ref{fig:16kbin}): runs with and without an IMBH again become well
separated after $\sim 5 t_{rh}(0)$. As expected, primordial binaries
carry lighter particles toward the center of the cluster (e.g., a 
$0.6+0.2 M_{\odot}$ binary will sink to the center like a 
$0.8 M_{\odot}$ single star, but will be observed as a single unresolved
source with the approximate luminosity of a $0.6 M_{\odot}$ star). 
Hence, mass segregation is partially suppressed when compared to the runs
where $f_b = 0$.  This difference in $\Delta\langle m \rangle $ is more
significant in the runs with a Miller \& Scalo IMF as compared to runs
with a Salpeter IMF. Because the runs drawn from a Miller \& Scalo IMF
lack massive remnants, binary stars become more gravitationally
dominant, and therefore have a more significant impact on the
dynamics. Fortunately, the binary-driven quenching of mass
segregation is weak when compared to IMBH-driven quenching and
thus it is possible to discriminate between systems with and without
an IMBH on the basis of $\Delta\langle m \rangle $, without the need of
assuming a binary fraction.

Combining the data from all our simulations with and without binaries,
we can identify three regions for the value of $\Delta\langle m \rangle $ in
a collisionally relaxed GC, irrespective of its binary
fraction:
\begin{itemize}
\item $\Delta\langle m \rangle \gtrsim0.1 M_{\sun}$. The system is unlikely to contain a central IMBH.
\item $\Delta\langle m \rangle \lesssim0.07 M_{\sun}$. The system is a good candidate to
harbor an IMBH.
\item $0.07 M_{\sun} \lesssim \Delta\langle m \rangle  \lesssim 0.1 M_{\sun}$. The system may or may
not contain an IMBH, depending on its binary fraction and on the
global IMF (and in particular on the number of massive dark remnants).
\end{itemize}

In addition, an estimate of the binary fraction based on the presence
of a parallel main sequence in the color-magnitude diagram is possible
for many observed clusters \citep{mil08}. Application of the mass
segregation diagnostic therefore can account for the actual number of 
binaries, resulting in a further reduction in the size of the region
of uncertainty.

Including the set of runs with $N=8192$, not shown in the plots but
whose $\Delta\langle m \rangle $ is reported in Tab.~\ref{tab:sim}, we see
no trends in $\Delta\langle m \rangle $ caused by an increase in the number
of objects in a cluster up to $N=32768$.  In
addition, the $N=131072$ control snapshots are consistent with
our results, strengthening the independence in the evolution of
$\Delta\langle m \rangle $.  An increase in the number of
particles reduces the deviation from snapshot to snapshot.
This actually improves the application of this indicator to actual GCs, 
where the number of stars is significantly larger than in our runs. 
Similarly, we see no significant trends in $\Delta \langle m \rangle$ 
caused by increasing the IMBH mass up to $M_{IMBH} = 0.03$, that is 3\% 
of the entire cluster (see Tab.~\ref{tab:sim}).  This suggests that 
reducing $M_{IMBH}$ below $0.01$ would still result in a 
quenching of mass segregation.

\subsection{The origin of IMBH-induced quenching of mass segregation}

The onset of mass segregation along with the initial contraction of 
the cluster brings the most massive stars and remnants into a dense environment.  
Even in clusters with only single stars, the dynamical formation of binaries
is inevitable.  Because $M_{IMBH}$ is much larger than the 
typical stellar mass, the IMBH has an 
extremely high probability of exchanging into a binary in a close 3-body encounter.
It therefore 
spends much of its lifetime in a binary or stable higher-N system (in more than 90\% of our snapshots the IMBH is a member of a multiple system).  
As a result, when massive main sequence stars in our simulations sink to the core 
after energy exchanges with other stars, they are efficiently ``heated
up'' and scattered away from the core in encounters with the IMBH and any companions it has.   
The IMBH stochastically moves around the core as a result of
these encounters and this further enhances the interaction rate
because the scatter cone is continuously replenished. This mechanism
for quenching mass segregation naturally explains the lack of
dependence of $\Delta \langle m \rangle $ on the number of particles used
and the minimal dependence on $M_{IMBH}$, as well as suggesting an additional 
explanation as to why the presence of primordial binaries further reduces 
mass segregation.

\section{Discussion and Conclusions}\label{sec:conclusion}

We have carried out a large set of direct N-body
simulations of star clusters with and without an IMBH including a
realistic mass spectrum and primordial binaries.  While previous
research has focused its attention mainly on the effects of an IMBH on
the surface brightness and velocity dispersion profiles of the
clusters --- signatures that are difficult to observe --- we searched
instead for a different fingerprint of the
presence of an IMBH. The existence of a massive, central object
quenches mass segregation and this effect manifests itself in
collisionally relaxed clusters through decreased radial variation in
the average mass of main sequence stars. This effect does not depend
on the mass of the black hole as long as it is dominant over the
typical mass of a star, nor on the details of the initial
configuration of the system such as initial mass function, density
profile and tidal field strength. The amount of mass segregation is
only weakly dependent on the binary fraction of the cluster. This
result allows us to use the amount of mass segregation to
separate collisionally relaxed clusters with and without an IMBH
without the need of additional modeling assumptions.

A critical requirement for the proposed signature is that the system 
be well-relaxed, so
that it has already attained equilibrium with respect to mass
segregation. From our simulations it turns out that this takes about
$5 t_{rh}(0)$. However we can only observe the current half-mass
relaxation time and this might be shorter than its initial value if
the system has lost a large fraction of its original mass. To compare
our simulations to observations, we must thus conservatively restrict
ourselves to GCs that:
\begin{enumerate}

\item Are not too influenced by the galactic tidal field (that is, with
a tidal to half-light radius $r_t/r_{hl} \gtrsim 10$, which corresponds
to tidal fields weaker than the weakest field in our simulations).

\item Have half-mass (3D) relaxation times below $\approx 1.5$~Gyr,
i.e. an age above $8 t_{rh}$. This leaves room for a mass loss of
about $50\%$ of the initial mass while still giving an integrated age of
about $5 t_{rh}$. In terms of observable quantities this translates into
a half-light relaxation time below $\approx 1$~Gyr. 

\end{enumerate}
Based on the \citet{har96} catalog, 31 galactic GCs
satisfy these stringent requirements in terms of relaxation time and
$r_t/r_{hl}$. The proposed diagnostic could probably be applied to more
clusters after properly evaluating a dynamical model for their
configuration and eventually accepting some uncertainty in the
selection of likely candidates to harbor an IMBH.

Thanks to the HST treasury survey of galactic GCs, data exist for the
cores of many clusters that explore deep enough to see main sequence
stars down to around $0.2 M_{\odot}$.  Along the same lines,
\citet{dem07}, among others, have also acquired images of clusters
around the half-light radius, in order to calculate the global
mass function of the system. The existing data from \citet{dem07} are
sufficient to apply this diagnostic to a few actual clusters and the
results from such a comparison will be presented in a companion paper.

In closing, we stress again that while the amount of mass segregation
has been proven here to be a viable indicator for the presence of an
IMBH in simulated star clusters, we cannot use this method alone to
claim the detection of an IMBH. However, by combining the measure of
mass segregation with all other constraints from the velocity
dispersion and surface brightness profiles, we can select the clusters
that seem most likely to harbor an IMBH, while at the same time
excluding some others from further scrutiny. Once we have identified
those clusters that are most promising, future observations, such as
proper motion studies, can focus their efforts to secure a robust
detection.

\acknowledgements

We thank Enrico Vesperini for useful discussions
and suggestions, Holger Baumgardt for sharing some of his data with
us and the referee for a careful reading of the manuscript and for 
constructive suggestions. This work was partially supported by NASA 
grant HST-AR11284.


\clearpage

\begin{table}
\footnotesize
\caption{Summary of the N-body simulations.} 

\label{tab:sim}
\begin{center}
\begin{tabular}{cccccccccc}
Name &N & $W_0$ & IMF & $M_{IMBH}$/$M_{tot}$ &
$M_{IMBH}$/$M_{\odot}$ & $f_b$ & $\langle{\Delta\langle m \rangle }\rangle$ & 
${\Delta\langle m \rangle }_{min}$ & ${\Delta\langle m \rangle }_{max}$ \\
8ks     & 8192  & 7.0 & Sal  & N/A   & N/A   & 0  & 0.07 & 0.040 & 0.112 \\
8km     & 8192  & 7.0 & M\&S & N/A   & N/A   & 0  & 0.13 & 0.095 & 0.167 \\
8kbs5   & 8192  & 5.0 & Sal  & N/A   & N/A   &0.1 & 0.09 & 0.029 & 0.138 \\
8kbs    & 8192  & 7.0 & Sal  & N/A   & N/A   &0.1 & 0.10 & 0.071 & 0.143 \\
8kbs11  & 8192  &11.0 & Sal  & N/A   & N/A   &0.1 & 0.09 & 0.037 & 0.116 \\
8kbm    & 8192  & 7.0 & M\&S & N/A   & N/A   &0.1 & 0.09 & 0.048 & 0.130 \\
16ks    & 16384 & 7.0 & Sal  & N/A   & N/A   & 0  & 0.11 & 0.071 & 0.158 \\
16ks.1  & 16384 & 7.0 & Sal  & N/A   & N/A   & 0  & 0.14 & 0.112 & 0.191 \\
16km    & 16384 & 7.0 & M\&S & N/A   & N/A   & 0  & 0.14 & 0.112 & 0.174 \\
16kbs   & 16384 & 7.0 & Sal  & N/A   & N/A   &0.1 & 0.09 & 0.060 & 0.140 \\
16kbm   & 16384 & 7.0 & M\&S & N/A   & N/A   &0.1 & 0.10 & 0.067 & 0.127 \\
32km    & 32768 & 7.0 & M\&S & N/A   & N/A   & 0  & 0.14 & 0.108 & 0.161 \\
128kk.1 &131072 & 7.0 &Kroupa& N/A   & N/A   & 0  & 0.13*& N/A   & N/A   \\
\tableline
8ksBH   & 8193  & 7.0 & Sal  & 0.03  & 104.0 & 0  & 0.05 & 0.014 & 0.091 \\
8kmbh   & 8193  & 7.0 & M\&S & 0.01  & 42.1  & 0  & 0.07 & 0.036 & 0.121 \\
8kmBH   & 8193  & 7.0 & M\&S & 0.03  & 129.3 & 0  & 0.06 & 0.011 & 0.097 \\
8kbsBH  & 8193  & 7.0 & Sal  & 0.03  & 114.4 &0.1 & 0.04 &-0.014 & 0.080 \\
8kbmbh  & 8193  & 7.0 & M\&S & 0.015 & 69.5  &0.1 & 0.04 &-0.021 & 0.079 \\
16ksbh  & 16385 & 7.0 & Sal  & 0.015 & 103.1 & 0  & 0.05 & 0.023 & 0.090 \\
16ksbh.1& 16385 & 7.0 & Sal  & 0.015 & 60.9   & 0  & 0.06 & 0.013 & 0.118 \\
16kmbh  & 16385 & 7.0 & M\&S & 0.015 & 128.2 & 0  & 0.08 & 0.027 & 0.113 \\
16kbsbh & 16385 & 7.0 & Sal  & 0.01  & 113.4 &0.1 & 0.04 & 0.015 & 0.078 \\
16kbmbh & 16385 & 7.0 & M\&S & 0.01  & 141.0 &0.1 & 0.05 & 0.015 & 0.084 \\
32kmbh  & 32769 & 7.0 & M\&S & 0.01  & 240.0 & 0  & 0.07 & 0.051 & 0.101 \\
128kkbh.1a&131072 & 7.0 &Kroupa& 0.013 &1000.0 & 0  & 0.09*& N/A   & N/A   \\
128kkbh.1b&131072 & 7.0 &Kroupa& 0.009 &1000.0 & 0  & 0.06*& N/A   & N/A   \\
\tableline
\end{tabular}
\tablecomments{We calculated the
  average, maximum and minimum values for $\Delta\langle m \rangle $ (in solar mass units)
  between 5 and 12 relaxation times for each run. Starred
  values are not averages, but are from a single snapshot.  The name of each run indicates: 
\begin{itemize}
\item The number of stars in the simulation (8k, 16k, 32k or 128k)
\item The presence of primordial binaries (b if $f_b > 0$)
\item The IMF (m for Miller \& Scalo, s for Salpeter, and k for Kroupa)
\item The presence of an IMBH (absent for none, bh for small BH mass, and BH for larger - see also fifth column)
\item The value of $W_0$ if different from $7.0$
\item Control run with IMF lower cut-off at $0.1 M_{\odot}$ (.1 suffix). 
\end{itemize}
}

\end{center}
\end{table}


\clearpage

\end{document}